\documentclass[aps,prl,twocolumn,showpacs]{revtex4}
\usepackage{graphicx} 
\usepackage{dcolumn}  
\usepackage{bm}       
\usepackage{amsmath}
\usepackage{epsfig}

\def\ii{{\rm i}}        

\begin{document}
\title{Non-local effects in the plasmons of strongly interacting nanoparticles, dimers, and waveguides}
\author{F.~J.~Garc\'{\i}a de Abajo}
\email{jga@cfmac.csic.es} \affiliation{Instituto de \'Optica - CSIC,
Serrano 121, 28006 Madrid, Spain}

\date{\today}

\begin{abstract}
Non-local effects in the optical response of noble metals are shown
to produce significant blueshift and near-field quenching of
plasmons in nanoparticle dimers, nanoshells, and thin metal
waveguides. Compared with a local description relying on the use of
frequency-dependent dielectric functions, we predict resonance
shifts as large as 10\% and field-intensity reduction of an order of
magnitude at inter-particle distances or metal thicknesses below
2\,\AA. Our results are based upon the specular-reflection model
combined with a suitable non-local extension of measured local
dielectric functions. We present a roadmap to design plasmon
resonances in nanometer metallic elements with application to
optical antennas and improved photovoltaic, light-emitting, and
sensing devices.
\end{abstract}
\maketitle

Knowledge of the optical response of materials at the nanoscale has
become a pillar of nanophotonics. While excellent agreement between
theory and experiment is generally achieved for nanoparticles and
nanostructured materials by relying on local, frequency-dependent
dielectric functions \cite{BH98,paper115}, non-local effects are
known to play an important role at small distances in the few- or
sub-nanometer region \cite{KV95}.
This range of distances is becoming experimentally feasible in
metallic dimers \cite{DN07}, nanoshells \cite{PRH03}, and tips
\cite{paper140}, which exhibit large near-field enhancement
\cite{LSB03,paper114} with practical application to improved SERS
biosensing \cite{XBK99,paper140}, photovoltaics \cite{DLM06}, and
solid-state lightning \cite{MBA06}. However, no account of non-local
effects has been reported in such systems, due in part to the
complexity of non-local descriptions of the optical response in
inhomogeneous environments.

First-principles calculations of the optical response are only
available for relatively simple systems, such as bulk metals and
small clusters of noble metal atoms \cite{IWY07}. However, several
approximate prescriptions have been elaborated to deal with more
complicated systems, and most notably the $d-$function formalism of
Feibelman \cite{F82} and the specular reflection model (SRM)
\cite{RM1966,W1966,dispersion4}. The latter is particularly
advantageous because it permits expressing the response of bounded
homogenous media in terms of their non-local bulk dielectric
function, as already reported in studies of fast electrons and ions
moving near planar surfaces \cite{dispersion2}.

In this Letter we carry out SRM calculations of dimers, shells, and
thin waveguides made of gold and silver. We show that non-local
effects produce significant plasmon blueshifts and reduction of the
near-field enhancement. The response is constructed from measured
optical constants, which are extended to include non-local effects
in the contribution of valence electrons. The validity of this model
is corroborated by successfully explaining experimentally observed
quantum-confinement effects leading to broadening and blueshift of
plasmons in small metal particles. We draw conclusions on more
complicated systems of relevance in plasmonics by applying the same
model.

\begin{figure}[b]
\centerline{\includegraphics*[width=7.5cm]{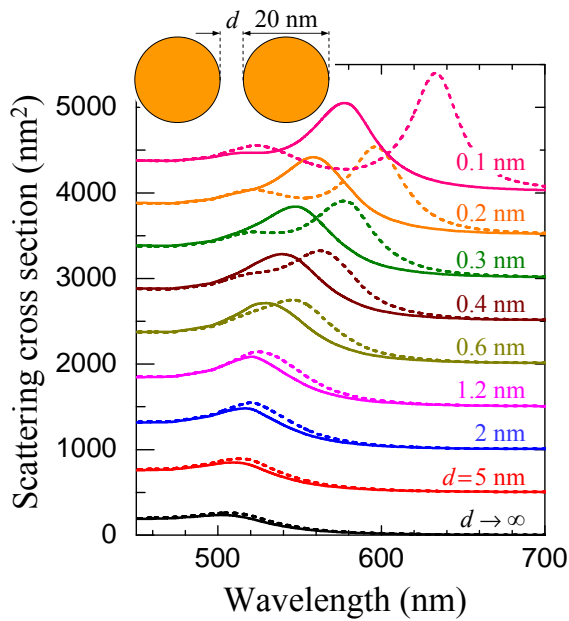}}
\caption{Extinction spectra of dimers formed by spherical gold
particles of 20-nm in diameter. Local (broken curves) and non-local
(solid curves) calculations are compared for several separations
between the particle surfaces, $d$. Consecutive curves are offset
vertically by 500\,nm$^2$ for clarity. The external electric field
is parallel to the inter-particle axis.} \label{Fig1}
\end{figure}

A first example of the blueshift produced by non-locality is shown
in Fig.\ \ref{Fig1} for scattering spectra of gold dimers calculated
with (solid curves) and without (dashed curves) inclusion of
non-local effects. The shift becomes more apparent as the
inter-particle distance is reduced and it is accompanied by sizeable
plasmon broadening. This lowest-order longitudinal dipole mode of
the dimer involves huge piling up of induced charge in the particle
junction region \cite{paper114}, thus enhancing the role of
non-locality.

In the SRM used throughout this work, each interface separating two
homogeneous media is described by surface charge distributions on
either side. These surface charges are used to construct {\it
boundary} fields as if they were surrounded by an infinite
homogeneous medium. The surface charges are determined by imposing
the continuity of the electric potential and the normal displacement
across the interface \cite{dispersion2}. The response of the
inhomogeneous system is thus reduced to the knowledge of non-local
bulk dielectric functions of the materials involved.

Homogeneous media are conveniently described in momentum space $q$,
in which non-locality shows up as a dependence of the dielectric
function $\epsilon(q,\omega)$ on $q$. In general insulators behave
as local media, whereas delocalized $s$ valence electrons in noble
metals constitute a source of non-locality. It is thus reasonable to
treat the response of both insulators and core-polarization in
metals as local and to approximate the valence-electron contribution
in the $q\rightarrow 0$ limit by a Drude formula,
$\epsilon^D(\omega)=1-\omega_p^2/\omega(\omega+\ii\eta)$, where
$\omega_p$ is the valence-electron-gas plasma frequency and $\eta$
is the plasmon width. Delocalized valence excitations produce a
significant $q$ dependence, as studied by Lindhard in the so-called
random-phase approximation (RPA) \cite{L1954}, and later by Mermin
to include electron-motion damping in a self-consistent fashion
\cite{M1970}. We adopt here the dielectric function proposed by
Mermin [$\epsilon^M(q,\omega)$] to describe the full non-local
contribution of valence electrons. This function has the desirable
$q\rightarrow 0$ limit $\epsilon^M(0,\omega)=\epsilon^D(\omega)$.
The resulting non-local permittivity of the metal is thus
approximated by \cite{dispersion3}
\begin{eqnarray}
\epsilon(q,\omega)=\epsilon^{\rm
loc}(\omega)-\epsilon^D(\omega)+\epsilon^M(q,\omega), \label{epsqw}
\end{eqnarray}
where $\epsilon^{\rm loc}(\omega)$ is the local, frequency-dependent
part of the response, taken from optical measurements \cite{JC1972}.
We correct the latter in Eq.\ (\ref{epsqw}) by subtracting the local
valence-electrons contribution $\epsilon^D$ and adding the non-local
valence-electrons response $\epsilon^M$. Finally, $\omega_p$ and
$\eta$ are chosen to fit both the long-wavelength tail and the
observed plasmon frequency of $\epsilon^{\rm loc}$
\cite{dispersion1} ($\hbar\omega_p=9$ eV and $\hbar\eta=0.05$ eV for
gold; $\hbar\omega_p=9.1$ eV and $\hbar\eta=0.02$ eV for silver).

\begin{figure}[ht]
\centerline{\includegraphics*[width=8cm]{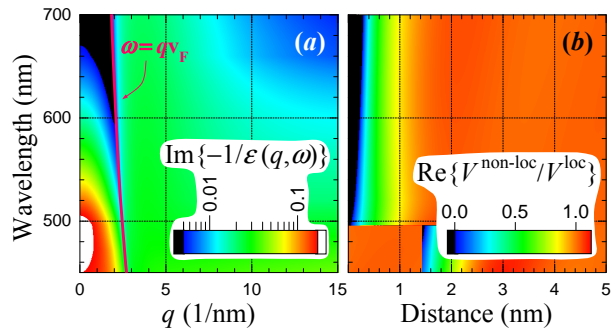}} \caption{{\bf
(a)} Non-local wavelength- and momentum-dependent response function
of homogeneous gold. {\bf (b)} Distance dependence of the non-local
interaction potential derived from (a) and normalized to the local
potential $1/[r\epsilon^{\rm loc}(\omega)]$.} \label{Fig2}
\end{figure}

The momentum and distance dependence of the response of bulk gold is
represented in Fig.\ \ref{Fig2}. In particular, Fig.\ \ref{Fig2}(a)
shows the so-called loss function of gold [${\rm
Im}\{-1/\epsilon(q,\omega)\}$] under the approximation of Eq.\
(\ref{epsqw}), which is directly accessible to electron energy-loss
spectroscopy \cite{BS1983}. Non-local effects become dramatic in the
optical region for $q\gtrsim 1$\,nm$^{-1}$ (i.e., to the left of the
Landau cutoff for excitation of valence electron-hole pairs,
$\omega=qv_F$, where $v_F$ is the Fermi velocity), and although this
is a large value compared to the momentum of light (e.g.,
$0.015$\,nm$^{-1}$ at 3\,eV), it is small compared to the inverse of
some relevant distances involved in plasmonic structures. Actually,
the real-space response $V^{\rm non-loc}$ obtained from the Fourier
transform of $4\pi/[q^2\epsilon(q,\omega)]$ differs considerably
from the local approximation $V^{\rm loc}=1/[r\epsilon^{\rm
loc}(\omega)]$ up to distances above $\sim 1$\,nm [Fig.\
\ref{Fig2}(b)]. The ratio between these two quantities has a real
part significantly smaller than 1 over that range of distances,
while the imaginary part lies below 0.2 in modulus within the area
explored in Fig.\ \ref{Fig2}(b).

\begin{figure}[hb]
\centerline{\includegraphics*[width=7cm]{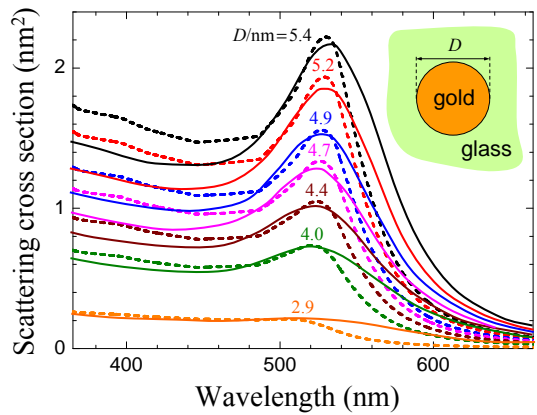}}
\caption{Extinction spectra of spherical gold particles embedded in
glass ($\epsilon=2.3$). The present non-local theory (dashed curves)
is compared to experiment (solid curves, taken from Ref.\
\cite{KV95}) for various values of the particle diameter (see text
insets). The calculated spectra have been shifted 15 nm to the red
to improve comparison.} \label{Fig3}
\end{figure}

It is useful to apply this formalism to spherical particles, the
optical response of which is conveniently described by their
multipolar response to external potentials of radial dependence
$r^l$, which develop an induced potential $-\alpha_l/r^{l+1}$, where
$l$ is the orbital momentum number and $\alpha_l$ is the
polarizability of order $l$. Applying the SRM, one finds
\cite{dispersion4}
\begin{eqnarray}
\alpha_l=a^{2l+1}\frac{l\epsilon_l-l\epsilon_h}{l\epsilon_l+(l+1)\epsilon_h},
\label{alphal}
\end{eqnarray}
where
\begin{eqnarray}
\frac{1}{\epsilon_l}=\frac{2a}{\pi(2l+1)}\int_0^\infty\frac{dq}{\epsilon(q,\omega)}\,j_l^2(qa),
\nonumber
\end{eqnarray}
$a$ is the particle radius, $\epsilon_h$ is the permittivity of the
local host medium surrounding the particle, and $j_l$ is the
spherical Bessel function of order $l$.

Fig.\ \ref{Fig3} shows good agreement because the present non-local
calculations (broken curves) and experimental data \cite{KV95}
(solid curves) for light scattering by small gold particles. Quite
different from the size-independent spectral profiles predicted by a
local description, non-local effects give rise to significant
broadening for particle radius below 2 nm [cf. Fig.\ \ref{Fig2}(b)].
These are the so-called quantum confinement effects, which show up
when the mean free path of valence electrons is comparable to the
particle diameter \cite{KV95}. This agreement between experiment and
theory supports our approach to describe non-local metal behavior at
short distances.

\begin{figure}
\centerline{\includegraphics*[width=7cm]{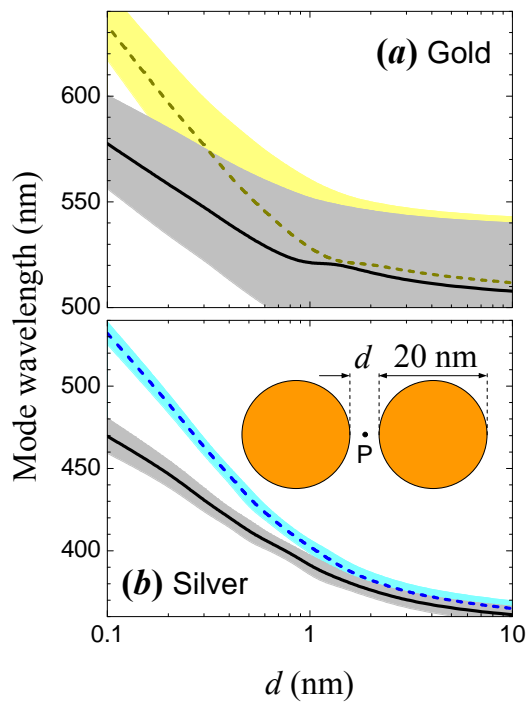}} \caption{Dimer
dipole-plasmon wavelength as a function of inter-particle surface
spacing $d$ for gold (a) and silver (b) 20-nm nanoparticles. Solid
and broken curves correspond to non-local and local descriptions of
the metal, respectively. The width of the modes is represented by
shaded regions covering the FWHM span of the
scattering-cross-section plasmon peaks.} \label{Fig4}
\end{figure}

For particle dimers [Fig.\ \ref{Fig1}] we describe each sphere by
the polarizabilities of Eq.\ (\ref{alphal}) and then perform
multiple scattering to obtain the self-consistent field
\cite{paper045}, including all multipoles with $l\le 20$ to achieve
convergence. The dependence of the lowest-order dipole-mode
wavelength on inter-particle separation in gold and silver dimers is
shown in Fig.\ \ref{Fig4} with (solid curves) and without (dashed
curves) inclusion of non-local effects. Besides significant
blueshifts, non-locality increases the plasmon broadening (shaded
areas), an effect that we attribute to the availability of
additional loss channels (valence electron-hole pairs) in the
metallic response.

\begin{figure}
\centerline{\includegraphics*[width=7cm]{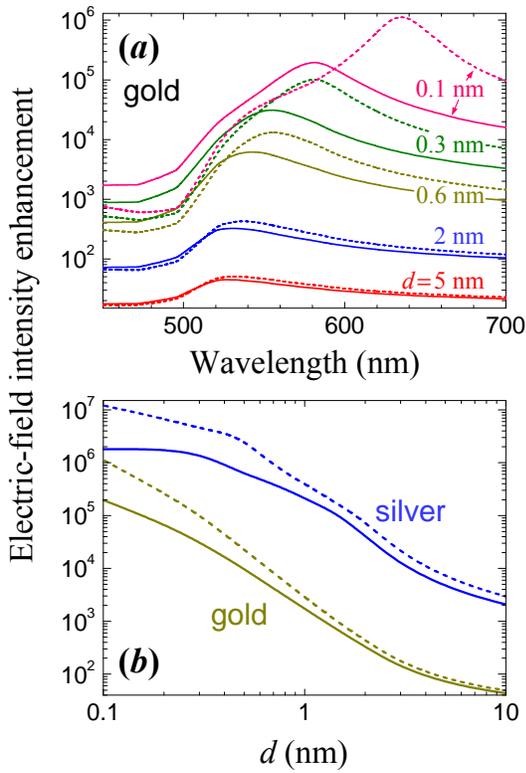}}
\caption{Enhancement of the electric field intensity relative to the
incident field ($|E/E^{\rm ext}|^2$) at the center of metallic
dimers with the same dimensions as in Fig.\ \ref{Fig4} (see point P
in the inset of that figure). {\bf (a)} Spectral dependence of the
enhancement for various separations between the surfaces of 20-nm
spherical gold particles. {\bf (b)} Maximum field enhancement as a
function of separation for gold and silver dimers. The results from
local and non-local descriptions of the response are shown by solid
and dashed curves, respectively.} \label{Fig5}
\end{figure}

Accompanying the far-field phenomenology discussed above, the near
field is also influenced by non-local effects, which can reduce the
field enhancement at the center of a dimer by nearly an order of
magnitude, as shown in Fig.\ \ref{Fig5}. This limits the
applicability of metal structures to SERS and non-linear optics, as
compared with the expectations opened by the field enhancement
predicted in the local approximation. Again, losses originating in
electron-hole-pair excitations contribute to this reduction, and so
does the finite penetration of the induced charge towards the
interior of the metal by a distance $\sim v_F/\omega$
\cite{dispersion2} (see Fig.\ \ref{Fig2}). This is in contrast to
the vanishing penetration associated to local descriptions.

\begin{figure}
\centerline{\includegraphics*[width=7cm]{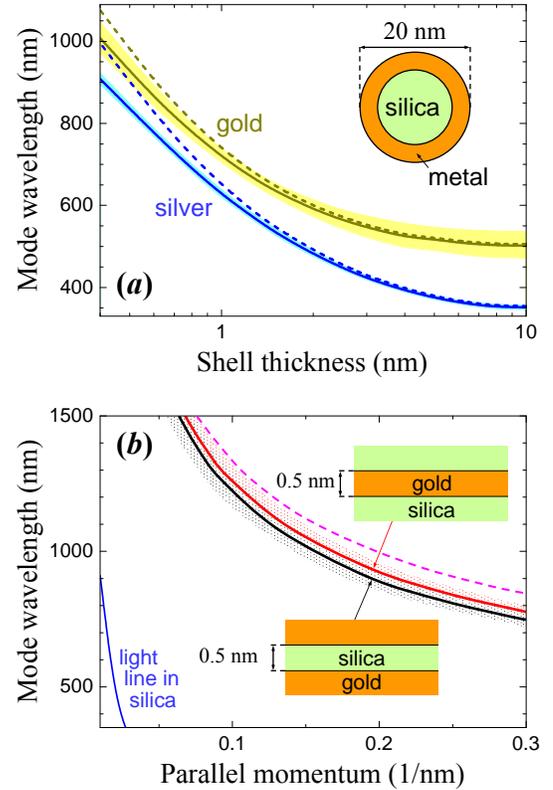}} \caption{{\bf
(a)} Dipolar mode wavelength of gold and silver shells surrounding a
silica core ($\epsilon=2.1$) as a function of metal thickness. Solid
and broken curves correspond to non-local and local descriptions of
the metal, respectively. The width of the modes is represented
through shaded regions covering the FWHM span of the
scattering-cross-section plasmon peaks. {\bf (b)} Plasmon dispersion
relations in gold-silica-gold and silica-gold-silica waveguides for
a 0.5-nm thick central layer (solid curves); a non-local description
yields the dashed curve in both cases.} \label{Fig6}
\end{figure}

Finally, non-local effects are not exclusive of small particles or
inter-particle gaps. They are also important in other
strongly-interacting metallic geometries typically encountered in
plasmonics like nanoshells [Fig.\ \ref{Fig6}(a)] and waveguides
[Fig.\ \ref{Fig6}(b)]. Similar to dimers, the plasmon modes in these
structures undergo bluedshifts exceeding 10\% at thicknesses below
5\,\AA. Interestingly, the mutually complementary geometries of the
dielectric gap and the metal film considered in Fig.\ \ref{Fig6}(b)
should yield the same dispersion relation in the non-retarded limit
(dashed curve), but non-local effects produce different dispersion
relations for each of these systems.

More elaborate methods have also been used to describe non-locality,
particularly using density-functional theory \cite{IWY07} and
plasmon quantum models \cite{PNH03}, although they become
impractical in large systems or non trivial geometries. In contrast,
the SRM used here accounts for the main features of non-locality and
is sufficiently versatile to deal with relatively complex systems
like those discussed above.


In summary, we have shown that non-local effects are relevant in the
response of metal systems involving short distances below a few
nanometers. In particular, we have reported large blueshifts and
reduced near-field intensity compared to a local description in
metallic dimers, nanoshells, and thin films. These effects should be
important in the design of plasmonic elements on the nanometer
scale, for which the formalism provided here provides a suitable and
versatile framework.

I would like to thank Javier Aizpurua, Jeremy Baumberg, Peter
Nordlander, Angel Rubio, and Mark Stockman for fruitful discussions.
This work was supported by the Spanish MEC (NAN2004-08843-C05-05 and
MAT2007-66050) and by the EU-FP6 (NMP4-2006-016881 "SPANS").


\end{document}